# Predictions regarding the supply of $^{99}$Mo and $^{99m}$Tc when NRU ceases production in 2018

C. K. Ross[1,2] and W. T. Diamond[3]

June 2015

[1] Retired NRC physicist, Ottawa, ON, Canada, carlkross@gmail.com

[3] Retired AECL physicist, Deep River, ON, Canada, diamond_w45@yahoo.ca

[2]Author to whom correspondence should be addressed

E-mail:  carlkross@gmail.com

**Summary**

The NRU reactor in Chalk River had been scheduled to stop producing medical isotopes by the end of 2016 but the Government of Canada recently announced that it will remain available to support isotope production until its operating license expires on 31 March, 2018. NRU has the capability of producing up to 80 % of the world's requirements for $^{99}$Mo but is presently producing less than 20 %. There are a number of initiatives underway, both within Canada and around the world, to find alternative ways of producing $^{99}$Mo or its daughter, $^{99m}$Tc. We examine the status of the main proposals and conclude that it will be challenging for any of them to meet the required demand by the end of 2016. An additional year should be enough time for some of the proposals to complete the development of manufacturing facilities and achieve regulatory approval. It is likely that these operators will have enough production capability to make up for the shortfall when the NRU operating license expires.

**INTRODUCTION**

The most widely used isotope for medical imaging is $^{99m}$Tc. World-wide, approximately 30 million doses are used each year and usage is growing at a rate of a few percent per year[1]. Each dose of $^{99m}$Tc is about 20 mCi[^1] and the current purchase price for a single dose is about $40. Most $^{99m}$Tc, which has a half-life of about 6 hours, is derived from the decay of the parent isotope, $^{99}$Mo. Because the half-life of $^{99}$Mo is about 66 hours, the time scale during which $^{99m}$Tc can be stored and shipped is extended by an order of magnitude compared to the direct production of $^{99m}$Tc.

Research reactors have been the main source of $^{99}$Mo for several decades[2]. The fission of $^{235}$U leads to many fission products, including $^{99}$Mo, which is produced with an efficiency of about 6 %. The $^{99}$Mo is extracted from the $^{235}$U targets, purified, converted to molybdate, $MoO_4^{2-}$, and loaded onto alumina columns. Each column is loaded into a lead-lined technetium generator, containing up to 10 Ci of $^{99}$Mo. As the $^{99}$Mo decays, it forms pertechnetate, $TcO_4^-$, which can be washed off the column using a saline solution. The generator, which is about the size of a large thermos, can be milked repeatedly over a period of two weeks to recover additional $TcO_4^-$, as illustrated in Figure 1.

Three reactors, NRU in Canada, HFR in the Netherlands and SAFARI-1 in South Africa, account for more than 80 % of the world's supply of $^{99}$Mo. The supply was stable for several decades but the unexpected shut-down of NRU in 2007 led to a shortage and what has come to be known as the isotope crisis[3]. It became generally recognized that the world's supply of its most important medical isotope relied on a few aging reactors whose future was uncertain because of licensing issues, technical challenges or government support.

A second challenge that affects most of the reactors over the longer term is the desire to limit the amount of highly-enriched uranium (HEU) in circulation. Although low-enriched uranium (LEU) can be used as fission targets for $^{99}$Mo production only two smaller producers (SAFARI-1 and OPAL) have successfully converted.

The isotope crisis led to an urgent effort to look for options to maintain a reliable supply of $^{99}$Mo. In the US, the National Nuclear Security Administration (NNSA) made funds available to support new proposals. In Canada, Natural Resources Canada (NRCan), which oversees the operation of the NRU reactor, called for proposals to look for alternative ways of producing

---

[^1]: The SI unit for activity is the Bq. However, the Ci (or mCi) is still widely used in nuclear medicine. One Ci is equal to 37 GBq.

$^{99}$Mo. In this article we will focus on those projects which are most likely to make a significant contribution to $^{99}$Mo or $^{99m}$Tc production when NRU shuts down in the spring of 2018.

**BASIC REQUIREMENTS**

A successful proposal to produce large quantities of $^{99}$Mo or $^{99m}$Tc on a regular basis must meet a number of requirements. The basic physics must be sound and used to show adequate production rates. Engineering challenges, such as heat dissipation, the handling of radioactive targets, the disposal of radioactive waste and ease of maintenance must be addressed. A suitable technique must be available for extracting the $^{99m}$Tc with high efficiency. Regulatory approval related to radiation safety and patient safety must be obtained. The cost per unit dose must be shown to be competitive with other proposals.

Options for producing $^{99}$Mo/$^{99m}$Tc can be divided into two broad categories. The first is based on neutron-induced fission of $^{235}$U. The fission route is exemplified by current reactor production either using HEU or LEU targets[2]. The fission of $^{235}$U can also be achieved using sub-critical assemblies which are driven by a particle accelerator that serves as a source of neutrons[4].

The second approach is based on nuclear reactions which lead to either $^{99}$Mo or $^{99m}$Tc directly. Although there are several choices, three reactions have received most attention. The first[5] is the neutron capture reaction by $^{98}$Mo, leading to $^{99}$Mo. Usually the $^{98}$Mo targets are irradiated in a nuclear reactor because the fluence rate from other neutron sources is not large enough to be relevant. The second[6] is the photoneutron reaction on $^{100}$Mo which can be written as $^{100}$Mo ($\gamma$, n) $^{99}$Mo. The threshold for this reaction is 9 MeV and the peak of the giant dipole resonance lies at about 14.5 MeV. A suitable photon beam can be generated using an electron linear accelerator (linac) and a bremsstrahlung radiator. The third reaction[7] is induced by a proton beam generated by a cyclotron striking a target of $^{100}$Mo. In this case, the reaction can be written as $^{100}$Mo (p, 2n) $^{99}$Tc and leads directly to both $^{99}$Tc and $^{99m}$Tc. It is the metastable state that is of interest for imaging, but the ground state component will compete with the metastable state when the isotope is tagged to biologically active molecules.

**PRODUCTION YIELDS**

An unusual unit, referred to as the six-day curie, is used presently for the sale and delivery of $^{99}$Mo. A six day curie is defined[2] as the amount of $^{99}$Mo activity left six days after the generator has left the producer's facilities. Six days represents more than two half-lives and suggests that only a little more than 20 % of the $^{99}$Mo produced in the reactor is available in the clinic. The global demand for $^{99}$Mo is estimated to be about 10,000 six-day curies per week[1] and that for Canada about 420 six-day curies per week[8].

The concept of the six-day curie may not be particularly relevant for some of the new modes of production and it has no relevance at all for $^{99m}$Tc produced directly by cyclotrons where $^{99}$Mo plays no role. Thus, it is worth considering production requirements from the known consumption rates of $^{99m}$Tc. There are about 5500 $^{99m}$Tc scans per day in Canada, each requiring about 20 mCi or a total of 110 Ci. $^{99m}$Tc decays with a half-life of about six hours so allowing for scans to be spread out over a working day would require about 220 Ci available at the beginning of each day. This amount of $^{99m}$Tc can be obtained by milking 280 Ci of $^{99}$Mo which would be the total inventory of $^{99}$Mo required. Over 24 hours, this inventory would decay by 60 Ci which is the amount of $^{99}$Mo that would need to be replaced on a daily basis. Allowing about a day for source preparation and shipping, would suggest that a national facility producing about 80 Ci of $^{99}$Mo per day, would meet Canada's requirements. This is more than a factor of three less than the present production rate. This estimate is consistent with work carried out at Idaho National Laboratories where they showed that a daily production rate of 25 Ci, if efficiently processed and delivered, could supply the state of Florida which has a population about half that of Canada[9].

For an arbitrary radioisotope, the activity per unit volume, $A$, after an irradiation time, $t$, is given by

$$A = \phi \cdot n \cdot \sigma \cdot (1 - e^{-\lambda t}), \tag{1}$$

where $\phi$ is the particle fluence rate, $n$ is the number of target atoms per unit volume and $\sigma$ is the cross section for producing an atom of the desired isotope. The product of these three terms gives the atomic production rate. The quantity in parentheses accounts for the decay of each atom where $\lambda$ is the decay constant. This quantity reaches a value close to unity after about five half-lives of irradiation (sometimes called the saturated yield) at which point the rate of production and of decay are equal.

The total production rate is obtained by integrating Eq. (1) over the volume of irradiated target atoms. The cross section for a given reaction is fixed so the production rate can only be changed by changing the fluence rate of the incident particles or the number of target atoms irradiated. The cross section for the production of $^{99}$Mo by $^{235}$U fission is about 36 b. This is to be contrasted with the maximum $^{100}$Mo (γ, n) $^{99}$Mo and $^{100}$Mo (p, 2n) $^{99}$Tc cross sections which are less than 0.25 b. Furthermore, the volume of target material that can be irradiated by either an electron accelerator or cyclotron is much smaller than that of a reactor.

Thus, a single 10 MW reactor, such as the MAPLE[2], is capable of producing the world's requirements of $^{99}$Mo/$^{99m}$Tc, while about 200 cyclotrons or 50 linacs would be required. Two 35 MeV, 50 kW linacs could supply all of Canada's present requirements for $^{99}$Mo.

**$^{99m}$TC EXTRACTION**

No matter what reaction is used to produce $^{99}$Mo or $^{99m}$Tc, a process must be available to extract efficiently the $^{99m}$Tc from the molybdenum matrix. In the case of fission, there is no molybdenum present to start with, so when the molybdenum is separated from the $^{235}$U targets, the $^{99}$Mo isotope is only diluted by whatever other molybdenum isotopes have been produced as fission products. The resulting material is said to have high specific activity, typically greater than 10,000 Ci/g. In this case, several curies of $^{99}$Mo can be bound to an alumina column a few centimeters in length, as described in the Introduction and illustrated in Figure 2. The resulting technetium generator is a simple, passive device that is easy to use and is presently the most widely used method of obtaining $^{99m}$Tc. An advantage of continuing to derive $^{99}$Mo as a fission product is that the existing technetium generators can be used, so that from the point of view of the user nothing changes.

The two nuclear reactions that are being proposed to produce $^{99}$Mo require starting with a molybdenum target. In these cases, only a small fraction (ppm) of the initial target is converted to $^{99}$Mo and the resulting material is said to have low specific activity, typically in the range of 1 to 10 Ci/g. The standard alumina column cannot be used in this case because the column would become impractically long. A $^{100}$Mo target is also used In the case of the (p, 2n) reaction that is used to produce $^{99m}$Tc directly. Several techniques have been developed over the years to separate $^{99m}$Tc from a molybdenum matrix[10]. We will briefly summarize the three that have received most attention. In all cases, the first step is to dissolve the molybdenum target and this can be done using a concentrated solution of hydrogen peroxide.

**Organic Solvent Extraction (MEK Process)**

In this technique, the aqueous solution containing the molybdenum and $^{99m}$Tc decay product is mixed with an immiscible organic solvent, methyl ethyl ketone (MEK). Technetium oxide is soluble in MEK but not molybdenum oxide. After agitation, the MEK will float to the surface of the aqueous solution and can be removed, carrying with it the $^{99m}$Tc. The MEK is evaporated leaving behind the technetium oxide. The cycle can be repeated several times as the $^{99}$Mo decays to $^{99m}$Tc. This approach has a long history so its strengths and weaknesses are well established. It is being used by the Winnipeg Health Sciences Centre[11] to process targets irradiated using the linac at the Canadian Light Source (CLS).

**Thermal Separation**

This approach exploits the different vapor pressures of oxides of molybdenum and technetium. $MoO_3$ melts at about 800 C and at this temperature the vapor pressure of $Tc_2O_7$ is about five orders of magnitude greater than that of $MoO_3$. If a flow of oxygen is maintained over the molten $MoO_3$, the $Tc_2O_7$ will be transported downstream and will condense when the temperature falls below 400 C. Idaho National Laboratories[9] refined the procedure and showed that good separation efficiency could be obtained. This is the technique that Best Cyclotron Systems is planning to use[12] to separate $^{99m}Tc$ from $^{100}Mo$ targets irradiated by proton beams.

**Chromatographic Column (ABEC)**

This technique is based on the use of two aqueous systems which are immiscible. If the technetium oxide is soluble in one of the phases but not the other, then these aqueous biphasic systems (ABS) can provide a route for separating technetium from molybdenum. Rogers et al[13] developed a technique for attaching the active component, polyethylene glycol, of one of the phases to a solid support so that, when the second liquid phase is present, ABS-like conditions are established. The column, loaded with the active component, is referred to as an aqueous biphasic extraction column, or ABEC. The technetium oxide is retained on the column and will be washed off when the mobile phase is changed to water. This approach has been developed and refined by NorthStar Medical Radioisotopes[14] and an animal study has shown that the product of the NorthStar ARSII separator meets or exceeds the specifications of that from a technetium generator[15].

**ACTIVE PROPOSALS**

There have been many proposals in the past few years for addressing the short-fall in the supply of $^{99}Mo$/$^{99m}Tc$. Some proponents advocate the construction of new reactors for the sole purpose of producing radioisotopes. Others require sophisticated accelerator technology for driving sub-critical assemblies. We restrict our considerations to those projects that have advanced well beyond the planning stage and have some likelihood of contributing to the supply of $^{99}Mo$/$^{99m}Tc$ within the next few years.

**Reactors**

The OPAL research reactor in Australia began operation in 2007. It is fueled with LEU and uses LEU targets for isotope production. It presently produces less than 5 % of the global requirements for $^{99}$Mo but a decision was taken in 2012 to greatly expand the production of medical isotopes and a new processing facility is now under construction[16]. It is scheduled to begin production in 2016 and be capable of producing up to 30 % of the world's requirements. They have also committed to a new facility for processing the radioactive waste and it is scheduled to begin operation in 2017. This is the only significant new source of fission $^{99}$Mo that will be available before NRU is shut down in March 2018.

**Neutron Capture**

The irradiation of $^{98}$Mo in a nuclear reactor has been used in the past to produce small quantities of $^{99}$Mo. NorthStar Medical Radioisotopes[5] is planning to make significant quantities of $^{99}$Mo for the US market by irradiating molybdenum targets in the Missouri Research Reactor (MURR). The irradiated targets will be processed by NorthStar to form molybdate, which will be shipped to participating nuclear pharmacies. Each of these will be equipped with the NorthStar RadioGenix separator which is based on ABEC column technology. Their proposal is well advanced and has been submitted to the FDA for regulatory approval. They project being able to produce modest amounts of $^{99}$Mo during 2015 and 50 % of US requirements by the end of 2016. This project faces the challenge of being based on a single aging reactor without any backup but the production technology is well established.

**Sub-Critical Assemblies**

This approach aims to retain the advantage of producing $^{99}$Mo with high specific activity while avoiding some of the challenges of operating a nuclear reactor. SHINE Medical Technologies[17] propose using neutron generators based on the D-T fusion reaction to inject neutrons into a aqueous solution of low-enriched uranium. Referred to as a Subcritical Hybrid Intense Neutron Emitter, or SHINE, the neutrons from the generator will be multiplied by a beryllium reflector and by fission neutrons but the solution will remain subcritical. Because the fission products are produced in an aqueous solution separation of the molybdenum isotopes will be simplified.

Although aiming to produce about half of US requirements for $^{99}$Mo, the start-up date for SHINE has been pushed back and is now planned for 2018. The neutron generator technology is the most demanding aspect of the project. By using a gaseous tritium target, the proponents aim to greatly lengthen the lifetime of traditional D-T neutron generators. However, it means they need a large tritium inventory and techniques for purifying the tritium gas as it is quickly diluted by the deuteron beam. In short, the proposed neutron generator technology has not yet been proven in an industrial setting.

**Cyclotrons**

Two Canadian consortia[18] as well as Best Cyclotron Systems[12] are working on the direct production of $^{99m}$Tc using cyclotrons and the reaction $^{100}$Mo(p,2n)$^{99m}$Tc. The TRIUMF group[19] recently announced the successful production of clinically useful quantities of $^{99m}$Tc. Because the half-life of $^{99m}$Tc is only six hours, in order for this approach to service a large distributed population, major changes in the entire model of production and distribution will be required. It will be much closer to the model of production of PET isotopes, requiring a local cyclotron and local target handling, including the recovery of used target material. To meet all of Canada's requirements for $^{99m}$Tc may require some cyclotrons dedicated to its production. It may be a challenge to mount this scale of effort by 2018, let alone any time during 2016. However, cyclotrons should provide some Canadian cities with a reliable source of $^{99m}$Tc as early as 2016 and will provide a good demonstration of the overall process.

**Electron Linear Accelerators**

Although several groups have used the ($\gamma$, n) reaction to produce small amounts of $^{99}$Mo in the past, Idaho National Laboratories[9] were the first to study the possibility of producing large quantities of $^{99}$Mo. Their work, which was carried out in the late 1990s, was not commercialized because it was assumed the MAPLE reactors would produce a robust supply of medical isotopes.

Both NorthStar in the US and the new facility at the Canadian Light Source Incorporated (CLS) in Canada plan to use electron linacs to produce $^{99}$Mo via the (γ,n) reaction. The CLS is collaborating with the Winnipeg Health Sciences Centre to qualify the $^{99}$Mo and to develop a separation unit for use in Canada. The objective is to meet most of the requirements of Manitoba and Saskatchewan by the end of 2016. An additional linac in a new facility will be required to meet a larger fraction of Canada's needs and this may be in place by 2018. NorthStar plans to install 16 linacs at one facility in Beloit, Wisconsin[5] that is reported to be capable of producing 50 % of the US requirements.

Industrial electron accelerator technology is well developed and extensively used for radiation processing. The additional challenges that arise when irradiating molybdenum targets are related to the beam window that must handle the pulsed, high-current beam and the removal of heat from the molybdenum target. Conventional water cooling is capable of removing the heat but may lead to complications due to water radiolysis. NorthStar is working with US national laboratories to establish target cooling using helium gas[20].

**ISOTOPICALLY ENRICHED MOLYBDENUM**

The neutron capture process requires $^{98}$Mo for which the natural abundance is about 24 %. The NorthStar/MURR proposal to produce up to 50 % of US demand by this process will require many kilograms of enriched $^{98}$Mo and routine reprocessing of target material. Natural molybdenum has been used in the past but $^{95}$Mo, with an abundance of 16 %, has a high neutron cross section that will reduce the neutron flux and there is not sufficient irradiation space (with high neutron flux) to irradiate the required volume of natural molybdenum to meet the production requirements. Because the irradiation of $^{98}$Mo in MURR is viewed as a short-term solution until the NorthStar electron accelerators begin operation, the long-term availability of enriched $^{98}$Mo is not a critical issue.

The ($\gamma$, n) and (p, 2n) reactions require $^{100}$Mo for which the natural abundance is less than 10 %. Thus, the availability of molybdenum enriched with $^{100}$Mo is critical because it gives a ten-fold increase in yield over what could be achieved with naturally occurring molybdenum. If all of Canada's requirements for $^{99}$Mo were to be met using linacs, an inventory of less than 2 kg of $^{100}$Mo would be required[6]. Presently, $^{100}$Mo can be produced using centrifuges in Zelenogorsk, Russia at a rate of about 30 kg per year[21]. Isotopic purity is not important for the ($\gamma$, n) reaction but is more critical for the (p, 2n) reaction because other isotopes of molybdenum will lead to unwanted radioisotopes that will increase the patient dose.

Because only a small fraction of $^{98}$Mo or $^{100}$Mo is converted to $^{99}$Mo during each irradiation and because the isotopically enriched material is expensive it will be necessary to recover and recycle the target material. Tests have shown that this can accomplished with modest effort and with losses less than 5 %. Thus, once an initial inventory is established, it will only be necessary to replace the material lost during each cycle. For Canada, this would be less than 1 kg of $^{100}$Mo per year.

There is presently only one supplier of isotopically enriched molybdenum so in the long term a more robust supply chain will need to be developed. Urenco (Netherlands) operates centrifuges for uranium enrichment and has produced small quantities of molybdenum in the past. Advanced Applied Physics Solutions[22] has shown that it is practical to magnetically separate relevant quantities of $^{98}$Mo and $^{100}$Mo. Although not as far advanced, a new approach based on laser isotope separation[23] has been suggested as a way to produce $^{100}$Mo. Once a clear market is established new suppliers of isotopically enriched molybdenum will likely emerge. It is also worth noting that this material can be stockpiled indefinitely.

**ECONOMICS**

The purchase price of $^{99m}$Tc in 2015 is about \$2/mCi or about \$40 for a typical scan and represents only about 15 % of the total cost of a SPECT procedure. Concern has been raised that much of the present cost of producing $^{99}$Mo in research reactors is subsidized, especially regarding the cost of waste management. The OECD has carried out an economic study[24] and show that, presently, the reactor cost represents less than 0.4 % of the final cost of a unit of $^{99m}$Tc. Under their worst-case full cost recovery model, this percentage needs to increase to about 3 %. Although this represents a very significant increase to the cost charged by the reactor operators it has only a small effect on the cost of $^{99m}$Tc to the end user.

Few of the new proposals have provided cost estimates but a report prepared for the American Association for the Advancement of Science[25] has attempted to use available data to estimate the cost of producing a dose of $^{99m}$Tc. They indicate that the production cost using electron accelerators will be about 35 % of that using reactors, while the cyclotron cost will be similar or slightly less than the reactor cost, depending on whether or not it is a dedicated or multipurpose cyclotron. We are not aware of cost estimates for the NorthStar/MURR or SHINE proposals.

Although there have been several reports suggesting that new production methods will lead to very substantial price increases for $^{99m}$Tc to the end user the available data does not indicate that this will be the case.

**PREDICTIONS**

Having reviewed the main scientific, engineering and economic issues regarding the production of $^{99}$Mo or $^{99m}$Tc, we will now attempt some predictions regarding the future supply.

$^{99m}$Tc is, by a large margin, the most widely used radioisotope for nuclear medicine procedures and there is strong support for its continued availability. Thus, a market for $^{99m}$Tc seems assured for the indefinite future.

Five research reactors have reliably produced most of the world's requirements of $^{99}$Mo for many years except during the NRU crisis of 2007 and when both NRU and HFR (Netherlands) were down for unscheduled repairs during 2010. Once NRU stops routine production of $^{99}$Mo, North America will not have a local reactor supplying the demand for $^{99m}$Tc for the first time in many years. Europe will still be in a state of flux as aging reactors are shut down (Osiris in France during 2015) or refurbished (BR2 in Belgium) and replacements are delayed. The only new reactor project that will likely be ready to produce a significant quantity of $^{99}$Mo in the next few years is the OPAL reactor in Australia. Their maximum production rate is estimated to be about 30 % of global requirements so not enough to replace NRU and Osiris production.

The SHINE project has been delayed because of financial and technical challenges and they are still awaiting regulatory approval to begin construction. It is not likely to contribute any $^{99}$Mo before NRU shuts down.

Within Canada, direct production of $^{99m}$Tc via cyclotrons may be approved and ready for use in the areas near the major players in this field (Vancouver, Edmonton, Sherbrooke and Montreal with modest quantities in London and Hamilton) by the end of 2016 although it would not be surprising if regulatory approval delays start-up.

The Canadian linac project will likely produce some supply in Canada (Manitoba and Saskatchewan) by the end of 2016 but it is unlikely that regulatory approval will be obtained in that time frame. It will require an additional linac in a new facility to increase that supply to cover most of Canada. Depending on financing, that may be available during 2018.

The NorthStar/MURR project, using the neutron capture reaction with $^{98}$Mo, is the new initiative most likely to be producing significant commercial quantities of $^{99}$Mo in the next few years. They have already produced more than 400 Ci of $^{99}$Mo in a test run and their separator is undergoing review to obtain the necessary approvals. Their goal is to be producing about half of the US requirements by the end of 2016.

It is likely that large areas of Canada and the US will still require fission $^{99}$Mo between now and 2018 and that will need to come from OPAL and will likely require support from NRU. The demands for NRU production could also come from reduced European production during that time with the closure of Osiris.

Looking beyond 2018, there will likely continue to be significant supplies of reactor-produced material for several years, perhaps approaching 50 % of demand. Although there is pressure to establish prices that reflect full cost recovery, there are indications that this will not greatly increase the cost to the end user[24]. One of the great strengths of $^{99}$Mo produced by the fission of $^{235}$U is the simplicity of the generator used to "milk" the $^{99}$Mo column and as long as these units are available they will likely find a market. However, it is generally recognized that significant scheduling conflicts arise when research reactors are also used to produce a commercial product. The significant waste stream from the fission targets, which will increase in size as reactors convert to the use of LEU targets, is another reason why reactor production may decrease as other methods come on line.

Reactor irradiation of $^{98}$Mo avoids the problem of waste from fission targets. However, the NorthStar/MURR approach is based on an aging reactor fueled with HEU and which has no backup. As recognized by NorthStar, this can be at best a short-term solution and it is being used until their linac facility is established.

Although it has been demonstrated that suitable cyclotrons exist to produce relevant quantities of $^{99m}$Tc, several facilities must be established in or near major population centres. Each of these will require staffing with highly qualified personnel and must establish manufacturing practices and quality systems related to the production of $^{99m}$Tc. If robust supplies of $^{99}$Mo are available, it seems unlikely that nuclear pharmacies will want to devote the effort to obtaining cyclotron-produced $^{99m}$Tc. This means that cyclotron beam time will not be taken away from producing the very short-lived isotopes that are important to PET imaging and cannot be produced any other way. Nevertheless, cyclotrons may be an important source of $^{99m}$Tc in the short term if other sources of $^{99}$Mo are significantly delayed.

We believe that using electron accelerators to irradiate $^{100}$Mo targets will turn out to be the best long term solution for producing $^{99}$Mo. Industrial linacs, such as that pictured in Figure 3, are widely used for radiation processing so the technology is mature. Although these machines are generally restricted to energies below 10 MeV to avoid activation, Mevex Corporation has shown that it is straightforward to add additional sections to achieve the optimum energy for $^{99}$Mo production. A single national or regional facility can host several linacs with the size of the facility easily scalable, depending on demand. No uranium of any kind is required and there is negligible radioactive waste. Regulatory licensing is much simpler than for a reactor and the cost is much less. The facility is committed to making a commercial product so there are no competing demands for beam time. The cost per unit dose of $^{99m}$Tc is estimated to be less than that produced by a reactor.

The main change from the point of view of the nuclear pharmacy is that that they will use different technology for separating the $^{99m}$Tc from the parent solution of $^{99}$Mo. As discussed earlier, several separation technologies exist and have been tested in laboratory settings but the approach based on the ABEC column and developed by NorthStar (Figure 4) is the closest to field deployment. Each nuclear pharmacy would have one (or more) of these units and would receive shipments of solution containing $^{99}$Mo from the NorthStar linac farm in Beloit, Wisconsin. When the solution has decayed beyond use, it would be returned to NorthStar for recycling.

The next two to three years will likely see some uncertainty in supply as different technologies are used to produce varying amounts of $^{99}$Mo or $^{99m}$Tc. Beyond that, it can be expected that NorthStar will be producing large quantities of $^{99}$Mo. If SHINE and some of the proposed reactor projects come on-line, it might be that there will be a significant excess manufacturing capability by 2020.

**ACKNOWLEDGEMENTS**

We would like to thank Peter Brown, Walter Davidson and Raphael Galea for helpful comments on the manuscript.

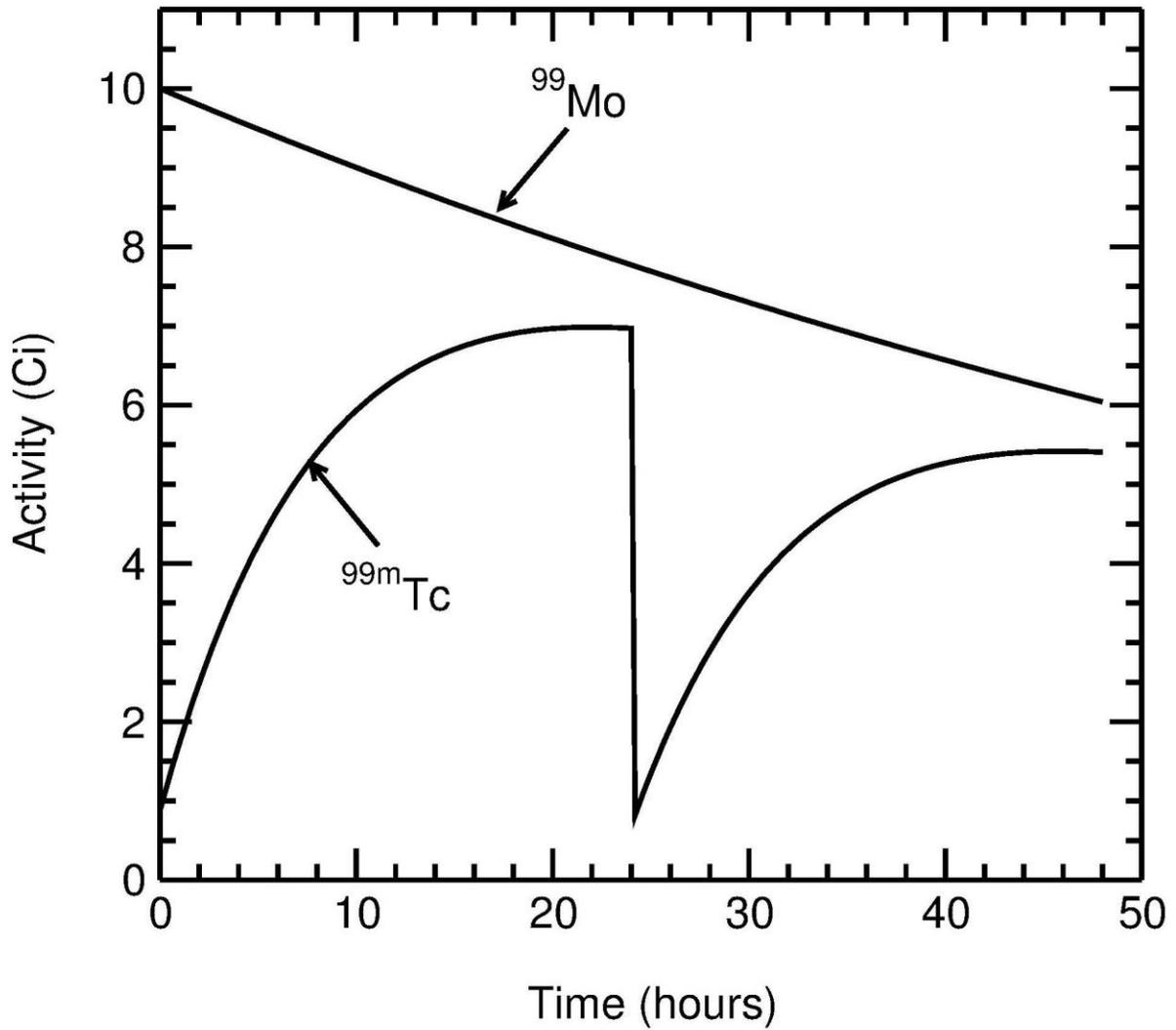

Figure 1. Performance of a typical technetium generator when milked every 24 hours. The decay of $^{99}$Mo with a half-life of 66 hours is shown by the upper curve. The generator was initially loaded with 10 Ci and has been milked at zero hours to remove the $^{99m}$Tc that has formed on the alumina column. $^{99}$Mo decays to $^{99m}$Tc over the next 24 hours at which time the column is milked again. This process can be repeated for about two weeks.

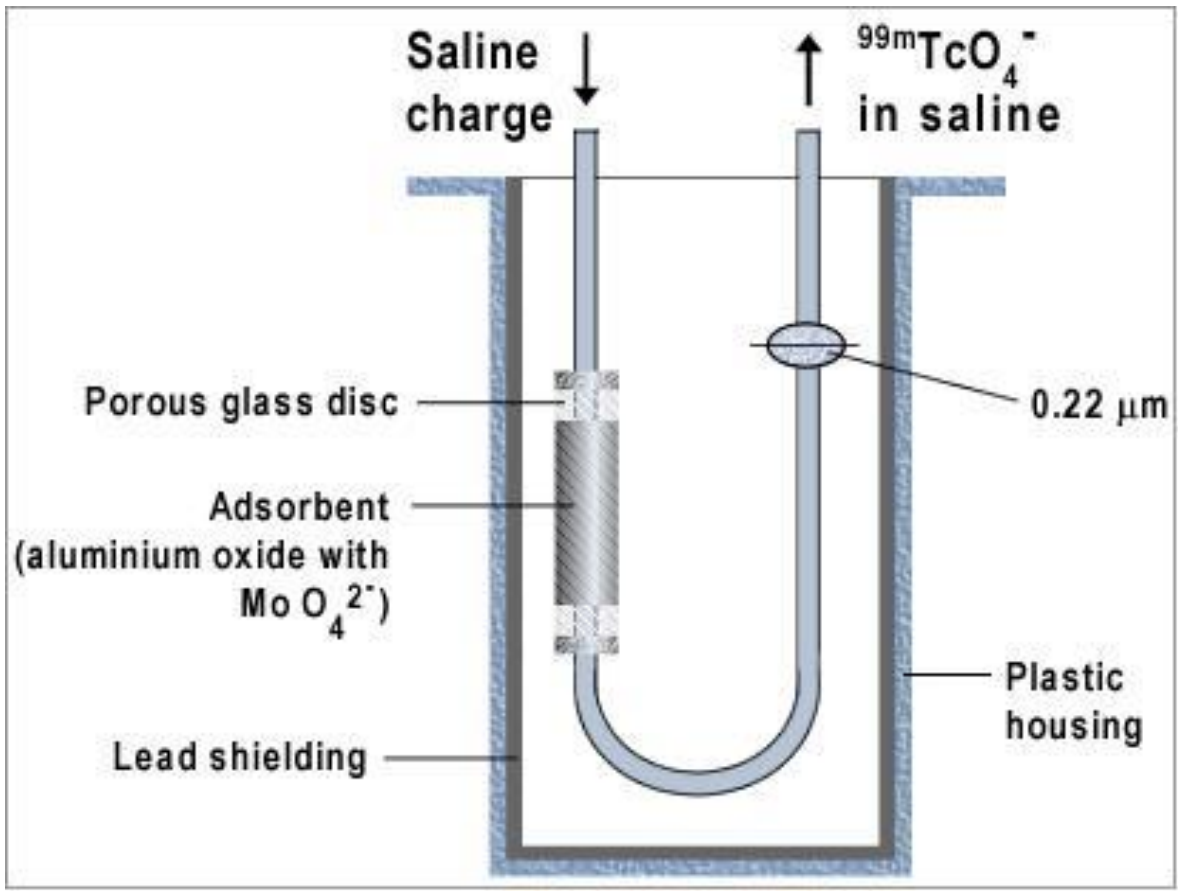

Figure 2. Typical components of a technetium generator. The molybdenum oxide is loaded on the alumina column by the manufacturer. The molybdenum decays to technetium and the resulting technetium oxide can be washed off the column using a salt solution. The process can be repeated for up to two weeks, by which time the molybdenum activity will be about 3 % of its original value. (Figure courtesy of the IAEA Human Health Campus).

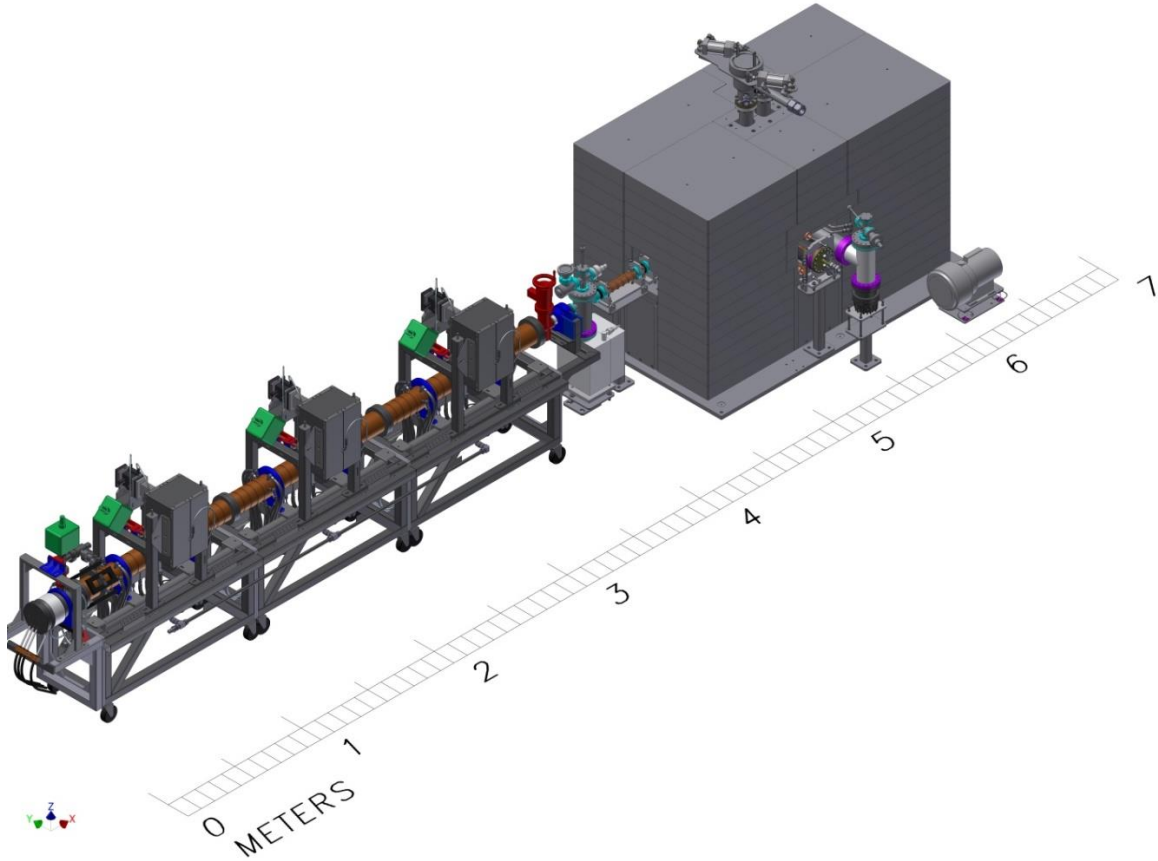

**Figure 3.** Drawing of the 35 MeV, 40 kW linac manufactured by Mevex Corporation (Stittsville, Ontario) and installed at CLS. The molybdenum target station is to the right, at the end of the machine and the complete assembly is about 6.5 m long. The molybdenum target is comprised of a series of discs, each about 1 cm in diameter. The shielding around the target assembly serves to localize induced activity, thus permitting assess to the room shortly after the end of irradiation. The three modulators to supply the RF power to the waveguides are installed in a separate area (Drawing courtesy of the Canadian Light Source).

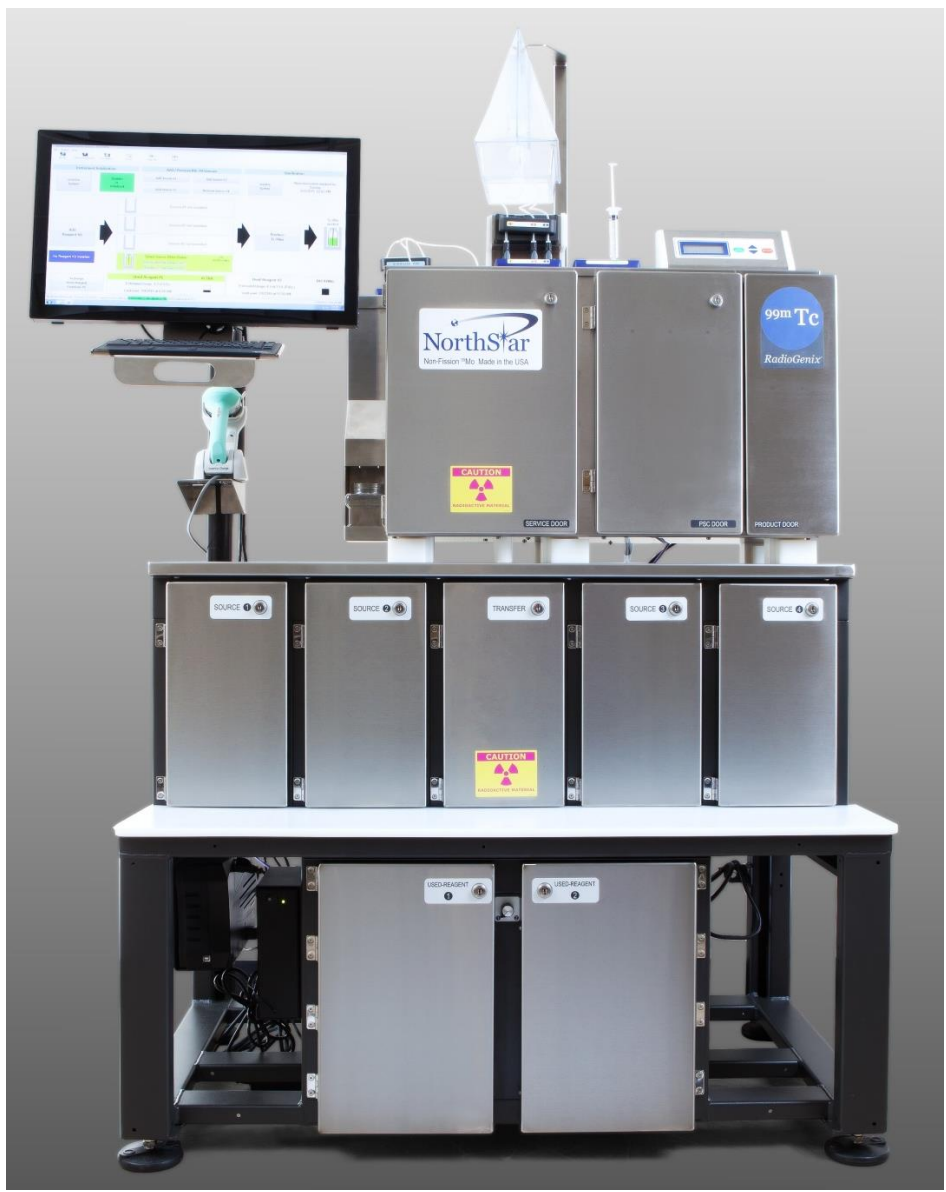

**Figure 4.** RadoGenix separator developed by NorthStar for separating $^{99m}$Tc from a solution of $^{99}$Mo. The unit is fully automated, operates at room temperature, does not require a fume hood and has a separation efficiency greater than 90 %. The required reagents are held in the clamshell structure at the top; the ABEC separation column is in middle unit on the top shelf and the $^{99m}$Tc product is delivered to the module on the right; up to four $^{99}$Mo sources can be loaded on the middle shelf; reagent waste is stored on the bottom shelf. The unit is presently undergoing assessment for US FDA approval. (Photo courtesy of NorthStar Medical Radioisotopes).